\documentclass[aps,superscriptaddress,reprint]{revtex4-1}
\usepackage{amsmath}
\usepackage{amssymb}
\usepackage{graphicx}
\usepackage[colorlinks,citecolor=blue, linkcolor=blue,hyperindex,CJKbookmarks,dvipdfm]{hyperref}

\begin{document}

\title{Nonreciprocity via nonlinearity and synthetic magnetism}
\author{Xun-Wei Xu}
\email{davidxu0816@163.com}
\affiliation{Department of Applied Physics, East China Jiaotong University, Nanchang,
330013, China}
\author{Yong Li}
\email{liyong@csrc.ac.cn}
\affiliation{Beijing Computational Science Research Center, Beijing 100193, China}
\author{Baijun Li}
\affiliation{Key Laboratory of Low-Dimensional Quantum Structures and Quantum Control of
Ministry of Education, Department of Physics and Synergetic Innovation
Center for Quantum Effects and Applications, Hunan Normal University,
Changsha 410081, China}
\author{Hui Jing}
\email{jinghui73@foxmail.com}
\affiliation{Key Laboratory of Low-Dimensional Quantum Structures and Quantum Control of
Ministry of Education, Department of Physics and Synergetic Innovation
Center for Quantum Effects and Applications, Hunan Normal University,
Changsha 410081, China}
\author{Ai-Xi Chen}
\email{aixichen@zstu.edu.cn}
\affiliation{Department of Physics, Zhejiang Sci-Tech University, Hangzhou
310018, China}
\affiliation{Department of Applied Physics, East China Jiaotong University, Nanchang,
330013, China}
\date{\today }

\begin{abstract}
We propose how to realize nonreciprocity for a weak input optical field via nonlinearity and synthetic magnetism.
We show that the photons transmitting from a linear cavity to a nonlinear cavity (i.e., an asymmetric nonlinear optical molecule) exhibit nonreciprocal photon blockade but no clear nonreciprocal transmission.
Both nonreciprocal transmission and nonreciprocal photon blockade can be observed, when one or two auxiliary modes are coupled to the asymmetric nonlinear optical molecule to generate an artificial magnetic field.
Similar method can be used to create and manipulate nonreciprocal transmission and nonreciprocal photon blockade for photons bi-directionally transport in a symmetric nonlinear optical molecule.
Additionally, a photon circulator with nonreciprocal photon blockade is designed based on nonlinearity and synthetic magnetism.
The combination of nonlinearity and synthetic magnetism provides us an effective way towards the realization of quantum nonreciprocal devices, e.g., nonreciprocal single-photon sources and single-photon diodes.
\end{abstract}

\maketitle

\section{Introduction}

Optical nonreciprocal transmission is an interesting phenomenon that photons transmit with asymmetric transmission coefficients under exchange of source and detector, and the nonreciprocal devices can be used to protect systems or devices from unwanted signals or noise~\cite{JalasNPT13}.
The conventional optical nonreciprocal devices are proposed by employing magneto-optical effects (Faraday rotation) to break reciprocity, and they have many disadvantages, such as they are bulky and work under large magnetic fields, so that it is a great challenge to implement them on chip.
To overcome all these drawbacks, various schemes have been developed to break optical reciprocity without the use of magneto-optical effects, including strategies based on nonlinearity~\cite{SoljacicOL03,ManipatruniPRL09,LFanSci12,QTCaoPRL17,HZShenPRA14} and synthetic magnetism~\cite{UmucalilarPRA11,KFangNPo12,TzuangNPt14,XuXWPRA17a,XuXWPRA17b,YYangSci19}.

In recent years, many works have reported the construction of optical isolators in an asymmetric nonlinear optical molecule, which consists of two coupled cavities with one of them containing nonlinear interactions, such as Kerr nonlinear interaction~\cite{MascarenhasEPL14,RodriguezPRA19}, optomechanical interaction~\cite{ZWangSR15,XuXWPRA18,LNSongArx19}, and the interaction to a two-level quantum emitter~\cite{ASZhengSR17} or two-level atomic ensemble~\cite{LNSongOC18}.
To demonstrate nonreciprocity in these isolators, the amplitude of the input field is usually pretty large, and such isolators are constrained by a dynamic reciprocity relation for the input fields with small amplitude~\cite{YShiNPt15}.
However, the reciprocity or nonreciprocity so far are defined based on the transmission coefficients, and there is non discussion on the statistical properties of the photons transport through an asymmetric nonlinear optical molecule.
Here, we will show that the transport photons exhibit strong photon antibunching in one direction but weak photon antibunching in the reverse direction, when a weak coherent field is injected into an asymmetric nonlinear optical molecule.
Such quantum nonreciprocal effect for transport photons with asymmetric statistical properties under exchange of source and detector is called nonreciprocal photon blockade, which is predicted firstly in a spinning Kerr resonator~\cite{RHuangarX18} based on the Fizeau light-dragging effect~\cite%
{MalykinPU00,HLv17,HJingOpt18,MaayaniNat18}. Nonreciprocal photon blockades were also predicted in multi-mode optomechanical systems based on directional nonlinear interaction~\cite{XuXWArx18} or quantum interference~\cite{BLiPR19}.
Nonreciprocal photon blockade may have important applications in the development of quantum nonreciprocal devices, which are crucial elements in chiral quantum technologies or topological photonics.

Different from the nonreciprocity induced by nonlinearity, the nonreciprocal transmission based on synthetic magnetism can work for weak signals and has been widely studied in a variety of systems, such as microwave
resonators connected with Josephson junctions~\cite{KochPRA10,SliwaPRX15,YPWangSR15,RoushanNPy16}, optomechanical systems with parametric coupling between optical and mechanical modes~\cite{MetelmannPRX15,XuXWPRA15,XuXWPRA16,SchmidtOpt15,XBYanFP19,KFangNPy17,BernierNC17,PetersonPRX17,BarzanjehNC17,YChenArX19}.
However, nonreciprocal photon blockade has not been considered in the systems with synthetic magnetism yet.

By introducing auxiliary modes to an asymmetric nonlinear optical molecule, we propose a class of structures containing both nonlinearity and synthetic magnetism, and show that these structures can be used to achieve both nonreciprocal transmission and nonreciprocal photon blockade. In this case, the photons transport one by one with high transmission coefficients in one direction but transport in pairs with low transmission coefficients in the reverse direction, which can be applied in quantum information processing as nonreciprocal single-photon sources or single-photon diodes.
The combination of synthetic magnetism and nonlinearity provides us an effective way to observe both nonreciprocal transmission and nonreciprocal photon blockade bi-directionally in a symmetric nonlinear optical molecule (both two coupled cavities are filled with nonlinear interactions), and design circulator with nonreciprocal photon blockade in a symmetric nonlinear cyclic three-mode system (all three mutually coupled cavities are filled with nonlinear interactions).

The remainder of this paper is organized as follows. In Sec.~II, we
study the transmission and statistical properties of photons transmitting in an asymmetric nonlinear optical molecule, and show that the transport photons exhibit nonreciprocal photon blockade but no clear nonreciprocal transmission when the input field is weak. In Sec.~III, we consider a system where one auxiliary mode is coupled to the asymmetric nonlinear optical molecule and show that the nonreciprocal transmission and nonreciprocal photon blockade can be obtained simultaneously or not by tuning the synthetic magnetic flux through the structure.  In Sec.~IV, the nonreciprocal transmission and nonreciprocal photon blockade are predicted when an auxiliary mode served as engineered reservoir is coupled to the asymmetric nonlinear optical molecule.
In Sec.~V, we show that nonreciprocal transmission and nonreciprocal photon blockade can appear bi-directionally when the asymmetric nonlinear optical molecule is replaced by a symmetric nonlinear optical molecule.
A circulator with nonreciprocal photon blockade is designed based on the combination of synthetic magnetism and nonlinearity in a symmetric nonlinear cyclic three-mode system, as shown in Sec.~V. Finally, a summary is given in Sec.~VI.

\section{Nonreciprocity in coupled nonlinear-linear cavities}

\begin{figure}[tbp]
\includegraphics[bb=160 260 426 513, width=7 cm, clip]{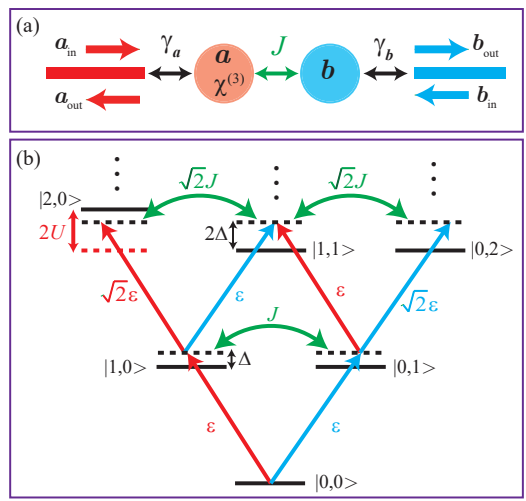}
\caption{(Color online) (a) Schematic diagram of an asymmetric nonlinear optical molecule consisting of a nonlinear cavity $a$ and a linear cavity $b$ with coupling strength $J$. (b) The
energy-level diagram: $|n_a,n_b\rangle$ (horizontal black short lines) represents the Fock state with $n_a$
photons in cavity $a$ and $n_b$ photons in cavity $b$, the red (blue) lines with
arrow on one end denote the excitations with photons injected from cavity $a$ (cavity $b$) and the green curves with
arrows on both ends denote the tunnel-coupling between the two cavities.}
\label{fig1}
\end{figure}

\begin{figure}[tbp]
\includegraphics[bb=199 311 396 509, width=8 cm, clip]{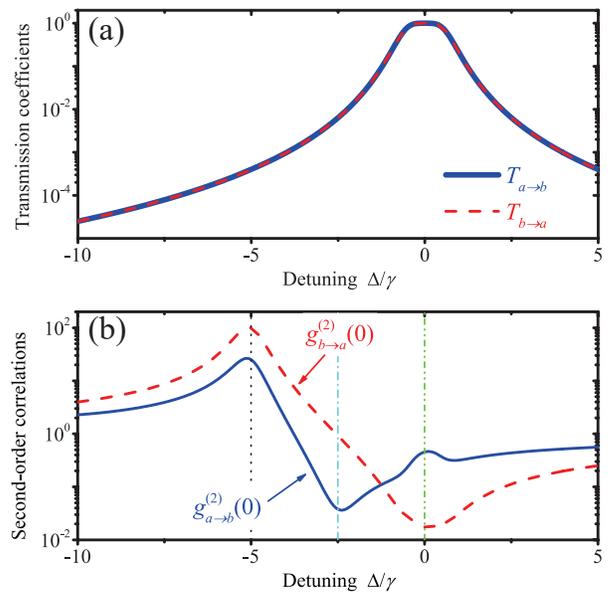}
\caption{(Color online) (a) The transmission coefficients $T_{a\rightarrow
b} $ (blue solid curve) and $T_{b\rightarrow a}$ (red dashed curve) are
plotted as functions of the detuning $\Delta/\protect\gamma$. (b) The
equal-time second-order correlation functions $g^{(2)}_{a%
\rightarrow b}(0)$ (blue solid curve) and $g^{(2)}_{b\rightarrow
a}(0)$ (red dashed curve) are plotted as functions of the detuning $\Delta/%
\protect\gamma$. Other used parameters are $\protect\gamma_a=\protect\gamma%
_b=\protect\gamma$, $\protect\varepsilon=0.01\protect\gamma$, $J=\protect%
\gamma/2$, and $U=5\protect\gamma$.}
\label{fig2}
\end{figure}

We consider an asymmetric nonlinear optical molecule consisting of an optical cavity ($a$, resonant
frequency $\omega _{a}$) filled with Kerr-type nonlinear material $\chi ^{(3)}$
(nonlinear interaction strength $U$) coupled to an empty optical cavity ($b$, resonant
frequency $\omega _{b}$) with tunnel-coupling strength $J$, as shown in Fig.~\ref{fig1}. The Hamiltonian
for the coupled nonlinear-linear cavities (asymmetric nonlinear optical molecule) is described by ($\hbar =1$)%
\begin{equation}
H_{\mathrm{ab}}=\omega _{a}a^{\dag }a+Ua^{\dag }a^{\dag }aa+\omega
_{b}b^{\dag }b+J\left( ab^{\dag }+a^{\dag }b\right) ,
\end{equation}%
under the rotating-wave approximation for $J\ll \min \left\{ \omega
_{a},\omega _{b}\right\} $, and $\gamma _{a}$ and $\gamma _{b}$ are the loss rates to the input or output field.
We assume that the system is working under the resonance $\omega _{a}=\omega _{b}$, impedance matching $J=\gamma _{a}/2=\gamma _{b}/2$, and strong nonlinear interaction $U\gg \max \left\{ \gamma
_{a},\gamma _{b}\right\}$ conditions.
In a microresonator, the Kerr interaction strength~\cite{Marin-PalomoNat17} is given by $U=\hbar \omega_a^2cn_2/(n^2V_{\rm eff})$, where $n_2$ ($n$) is the nonlinear (linear) refractive index, $c$ is the speed of light in vacuum, and $V_{\rm eff}$ is the nonlinear optical mode volume.
The integrated silicon nitride microresonator~\cite{Marin-PalomoNat17} and silica toroidal microresonator~\cite{QTCaoPRL17} provide us feasible ways to satisfy the strong nonlinear interaction $U\gg \max \left\{ \gamma_{a},\gamma _{b}\right\}$ conditions with ultrahigh-Q and small mode volume $V_{\rm eff}$.

In order to probe the transmission properties
and the statistical properties of the transport photons, we assume that a
weak laser (with amplitude $\varepsilon \ll \min{\{\gamma _{a},\gamma _{b}\}}$ and frequency $\omega _{p}$) is input from cavity $a$, or from cavity $b$, as%
\begin{equation}\label{Eq2}
H_{\mathrm{probe}}=\varepsilon \left( e^{-i\omega _{p}t}o^{\dag }+e^{i\omega
_{p}t}o\right) ,
\end{equation}%
for $o=a$ or $b$, with the detuning $\Delta\equiv \omega_a-\omega _{p}=\omega_b-\omega _{p}$.
According to the input-output relations~\cite{GardinerPRA85}, we have $a_{%
\mathrm{in}}=\varepsilon /\sqrt{\gamma _{a}}$ and $b_{\mathrm{out}}=\sqrt{%
\gamma _{b}}b$ for photons transmitted from cavity $a$ to cavity $b$, and $b_{\mathrm{in}}=\varepsilon /\sqrt {\gamma _{b}}$ and $a_{\mathrm{out}}=\sqrt{\gamma _{a}}a$ for photon transport from cavity $b$ to cavity $a$. Then the transmission coefficient
from cavity $a$ to cavity $b$ can be defined by
\begin{equation}
T_{a\rightarrow b}\equiv \frac{\langle b_{\mathrm{out}}^{\dag }b_{\mathrm{out%
}}\rangle }{\langle a_{\mathrm{in}}^{\dag }a_{\mathrm{in}}\rangle }=\frac{%
\gamma _{a}\gamma _{b}}{\varepsilon ^{2}}\left\langle b^{\dag }b\right\rangle,
\end{equation}
and the transmission coefficient from cavity $b$ to cavity $a$ can be defined by
\begin{equation}
T_{b\rightarrow a}\equiv \frac{\langle a_{\mathrm{out}}^{\dag }a_{\mathrm{out%
}}\rangle }{\langle b_{\mathrm{in}}^{\dag }b_{\mathrm{in}}\rangle }=\frac{%
\gamma _{a}\gamma _{b}}{\varepsilon ^{2}}\left\langle a^{\dag }a\right\rangle.
\end{equation}
Moreover, the statistic properties of the transmitted photons $b_{\mathrm{%
out}}$ and $a_{\mathrm{out}}$ can be described by the equal-time second-order
correlation functions in the steady state ($t\rightarrow \infty $), for photons transmitted from cavity $a$ to cavity $b$ as
\begin{equation}
g_{a\rightarrow b}^{(2)}(0)\equiv \frac{\langle b_{\mathrm{out}}^{\dag }b_{%
\mathrm{out}}^{\dag }b_{\mathrm{out}}b_{\mathrm{out}}\rangle }{\langle b_{%
\mathrm{out}}^{\dag }b_{\mathrm{out}}\rangle ^{2}}=\frac{\langle b^{\dag
}b^{\dag }bb\rangle }{\langle b^{\dag }b\rangle ^{2}},
\end{equation}%
and for photons transmitted from cavity $b$ to cavity $a$ as
\begin{equation}
g_{b\rightarrow a}^{(2)}(0)\equiv \frac{\langle a_{\mathrm{out}}^{\dag }a_{%
\mathrm{out}}^{\dag }a_{\mathrm{out}}a_{\mathrm{out}}\rangle }{\langle a_{%
\mathrm{out}}^{\dag }a_{\mathrm{out}}\rangle ^{2}}=\frac{\langle a^{\dag
}a^{\dag }aa\rangle }{\langle a^{\dag }a\rangle ^{2}}.
\end{equation}%

Both the transmission and statistical properties of the photons in the coupled nonlinear-linear cavities can be obtained by solving the master
equation~\cite{Carmichael93} for the density matrix $\rho$ of the system as
\begin{equation}
\frac{\partial \rho }{\partial t}=-i\left[H,\rho \right] +\gamma _{a}L[a]\rho +\gamma _{b}L[b]\rho ,
\end{equation}%
where $H=H_{\mathrm{ab}}+H_{\mathrm{probe}}$, $L[o]\rho =o\rho o^{\dag }-\left( o^{\dag }o\rho +\rho o^{\dag
}o\right) /2$ denotes a Lindbland term for an operator $o$. We assume that the frequencies of the
two optical modes are so high that the thermal effect can
be neglected.

The transmission coefficients $T_{a\rightarrow b} $ (blue solid curve) and $T_{b\rightarrow a}$ (red dashed curve)
under the weak driving condition $\varepsilon \ll \gamma_a=\gamma_b$ are shown as functions of the detuning $\Delta$ in Fig.~\ref{fig2}(a).
We have $T_{a\rightarrow b} \approx T_{b\rightarrow a}$, i.e., there is no clear nonreciprocal behavior in the transmission spectrum even with strong nonlinearity. In this case, the transmission spectrum relies primarily on the populations in the one-photon states, i.e., $|1,0\rangle$ and $|0,1\rangle$, which are nearly independent of the nonlinearity of the system for the populations in the two-photon states are much smaller than the ones in the one-photon states.

Different from the reciprocal transmission in Fig.~\ref{fig2}(a), there is a clear nonreciprocal behavior in the statistical properties of the transmitted photons as shown in Fig.~\ref{fig2}(b) with the same parameters. We have $g_{b\rightarrow a}^{(2)}(0) \neq g_{a\rightarrow b}^{(2)}(0)$, especially with the detuning around $\Delta=0$, which corresponds to the condition for maximum transmission $T_{a\rightarrow b} \approx T_{b\rightarrow a}\approx 1$. The strong photon blockade $g_{b\rightarrow a}^{(2)}(0)\ll 1$ for photons transmitted from cavity $b$ to cavity $a$ is induced by the large detuning for the transition from state $|1,1\rangle$ to $|2,0\rangle$ with the strong nonlinear interaction $U>\gamma_a=\gamma_b \sim J$ in cavity $a$, as shown in Fig.~\ref{fig1}(b). In contrast, when the photons are input from cavity $a$, the strong nonlinear interaction $U>\gamma_a=\gamma_b \sim J$ only blockades the transition $|1,0\rangle \nrightarrow |2,0\rangle \rightarrow|1,1\rangle \rightarrow|0,2\rangle$, but not the transition $|1,0\rangle\rightarrow |0,1\rangle \rightarrow|1,1\rangle \rightarrow|0,2\rangle$, so we have $g_{a\rightarrow b}^{(2)}(0) \approx 0.45$.

In addition, there are strong bunching effect [$g_{b\rightarrow a}^{(2)}(0)\gg 1$ and $g_{a\rightarrow b}^{(2)}(0)\gg 1$] for photons transmitted in both directions around the detuning $\Delta=U$, which is origin from the resonant two-photon excitation $|0,0\rangle \rightarrow |2,0\rangle$.
There is strong photon blockade $g_{a\rightarrow b}^{(2)}(0)\ll 1$ for photons transmitted from cavity $a$ to cavity $b$ around the point $\Delta=U/2$. This counterintuitive behavior can be understood by the fact that there is quantum interference between the two transitions $|1,0\rangle\rightarrow |2,0\rangle \rightarrow|1,1\rangle$ and $|1,0\rangle\rightarrow |0,1\rangle \rightarrow|1,1\rangle$, and the population in state $|1,1\rangle$ (as well as $|0,2\rangle$) are canceled out for destructive interference with $\Delta=U/2$.

\section{Nonreciprocity in cyclic three-mode systems}

\begin{figure}[tbp]
\includegraphics[bb=193 441 431 530, width=7 cm, clip]{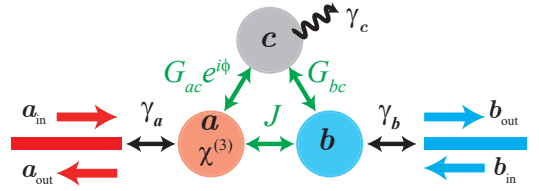}
\caption{(Color online) Schematic diagram of a cyclic three-mode system, i.e., an asymmetric nonlinear optical molecule (nonlinear mode $a$ and linear mode $b$) coupling to an auxiliary mode $c$ simultaneously, where $J$, $G_{ac}$ and $G_{bc}$ are the coupling strengths with the total phase $\protect\phi$.}
\label{fig3}
\end{figure}

In this section, we will show that the nonreciprocal behavior can be observed simultaneously or not in the transmission spectrum and the statistical properties of the transmitted photons, which can be controlled by a phase factor in a cyclic three-mode systems, as shown in Fig.~\ref{fig3}. Different from the model discussed in the last section, an auxiliary mode ($c$, with resonant frequency $\omega _{c}=\omega_a =\omega_b$) is added and coupled to modes $a$ and $b$ simultaneously. Without loss of generality, the coupling strengths $J$, $G_{ac}$, and $G_{bc}$ can be set as positive real numbers with the total phase $\phi$ obtained by redefining the annihilation operators ($a$, $b$ and $c$). It has been shown that the cyclic three-mode systems can be used to realize nonreciprocal transmission~\cite{XuXWPRA15}. Nevertheless, we show here that the nonreciprocal blockade can be observed in the cyclic three-mode system with Kerr-type nonlinear material $\chi ^{(3)}$ (nonlinear strength $U$) added in one mode.

The Hamiltonian of the cyclic three-mode system is given by $H=H^{(1)}_{\mathrm{abc}}+H_{\mathrm{probe}}$, with
\begin{eqnarray}\label{Eq8}
H^{(1)}_{\mathrm{abc}} &=& \omega _{a}a^{\dag }a+Ua^{\dag }a^{\dag }aa+\omega
_{b}b^{\dag }b+ \omega _{c}c^{\dag }c  \nonumber \\
&&+ \left( Jab^{\dag }+G_{bc}b c^{\dag }+G_{ac} e^{i \phi} c a^{\dag
}+\mathrm{H.c.}\right),
\end{eqnarray}
and $H_{\mathrm{probe}}$ in Eq.~(\ref{Eq2}).
As shown in Ref.~\cite{KochPRA10}, the total phase $\phi$ is formally equivalent to having a synthetic magnetic flux threading the plaquette formed by the three modes ($a$, $b$ and $c$). The time-reversal symmetry of the Hamiltonian is broken and  nonreciprocal transmission is observed for the synthetic magnetic flux $\phi \neq k\pi$ ($k$ is an integer).
In the numerical calculation of the transmission coefficients and second-order correlation functions, we set the damping rate of the added mode $c$ as $\gamma _{c}$, and the dissipation of the added mode $c$, i.e., $\gamma _{c} L[c]\rho$, is added to the master equation.

\begin{figure}[tbp]
\includegraphics[bb=102 244 484 556, width=8.5 cm, clip]{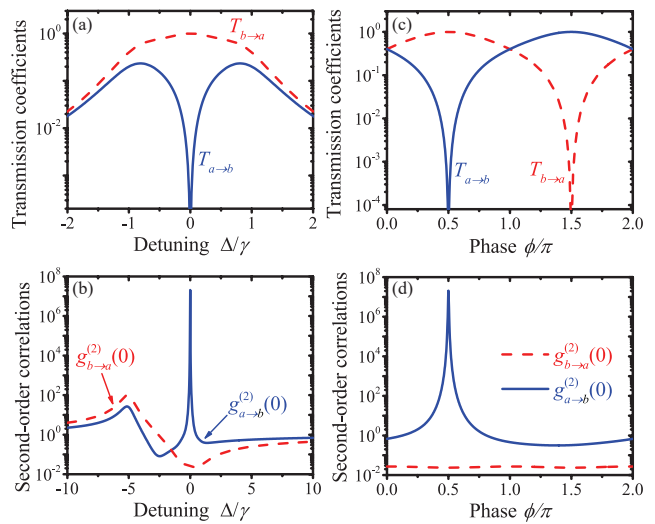}
\caption{(Color online) The transmission coefficients $T_{a\rightarrow
b} $ (blue solid curve) and $T_{b\rightarrow a}$ (red dashed curve) are
plotted as functions of the detuning $\Delta/\protect\gamma$ for $\phi=\pi/2$ (a) and the phase $\phi/\pi$ for $\Delta=0$ (c). The
equal-time second-order correlation functions $g^{(2)}_{a%
\rightarrow b}(0)$ (blue solid curve) and $g^{(2)}_{b\rightarrow
a}(0)$ (red dashed curve) are plotted as functions of the detuning $\Delta/\protect\gamma$ for $\phi=\pi/2$ (b) and the phase $\phi/\pi$ for $\Delta=0$ (d). Other used parameters are $\protect\gamma_a=\protect\gamma%
_b=\protect\gamma_c=\protect\gamma$, $\protect\varepsilon=0.01\protect\gamma$, $J=G_{ac}=G_{bc}=\protect%
\gamma/2$, and $U=5\protect\gamma$.}
\label{fig4}
\end{figure}

As expected, there is a high transmission from cavity $b$ to cavity $a$ with $T_{b\rightarrow a}\approx1$, but low transmission from cavity $a$ to cavity $b$ with $T_{a\rightarrow b}\approx 0$, with detuning $\Delta=0$ and phase $\phi=\pi/2$, as shown in Fig.~\ref{fig4}(a).
In the meantime, the photons transport from cavity $b$ to cavity $a$ exhibit strong antibunching effect, i.e., $g_{b\rightarrow a}^{(2)}(0)\ll 1$, and the photons transmitted in the versa direction exhibit strong bunching effect, i.e., $g_{a\rightarrow b}^{(2)}(0)\gg 1$ [see Fig.~\ref{fig4}(b)].
Physically, the strong antibunching effect, i.e., $g_{b\rightarrow a}^{(2)}(0)\ll 1$ is induced by the strong nonlinearity in cavity $a$, and the strong bunching effect, i.e., $g_{a\rightarrow b}^{(2)}(0)\gg 1$ originates from the population quenching of the one-photon state in cavity $b$ by destructive interference~\cite{XuXWPRA14} between the two paths for generating
one photon in cavity $b$, i.e., $a\rightarrow b$ and  $a\rightarrow c \rightarrow b$.
Therefore, we can obtain both the nonreciprocal transmission spectrum and the nonreciprocal blockade simultaneously via nonlinearity and synthetic magnetism in the cyclic three-mode system.

To demonstrate the nonreciprocity's tunability, the transmission coefficients $T_{b\rightarrow a}$ and $T_{a\rightarrow b}$ and the second-order correlation functions $g^{(2)}_{b\rightarrow a}(0)$ and $g^{(2)}_{a\rightarrow b}(0)$ are plotted as a function of the phase $\phi$ in Figs.~\ref{fig4}(c) and \ref{fig4}(d) when $\Delta=0$. The system shows reciprocal transmission $T_{b\rightarrow a}\approx T_{a\rightarrow b}$ and nonreciprocal blockade $g^{(2)}_{b\rightarrow a}(0)\ll g^{(2)}_{a\rightarrow b}(0)$ for phase $\phi=0$ or $\pi$. The optimal phase for nonreciprocal transmission ($T_{b\rightarrow a}\approx1$ and $T_{a\rightarrow b}\approx 0$) and nonreciprocal blockade [$g^{(2)}_{b\rightarrow a}(0)\ll 1$ and $g^{(2)}_{a\rightarrow b}(0)\gg 1$] is $\phi=\pi/2$.

\section{Nonreciprocity via reservoir engineering}

Actually, the approach of introducing nonlinearity in time-reversal symmetry broken system can be generalized to
a wide range of systems to observe both the nonreciprocal transmission and the nonreciprocal photon blockade.
A general method is proposed for generating nonreciprocal behavior in cavity-based
photonic devices by employing reservoir engineering in Ref.~\cite{MetelmannPRX15}.
As a simple example, the engineered reservoir can be obtained by adding an auxiliary mode, i.e., $c$, with large damping rate, i.e., $\gamma_c\gg\gamma_a=\gamma_b$.
In this large damping limit, the mode $c$ can be described as a general Markovian reservoir.
By eliminating the mode $c$ adiabatically (see Appendix~\ref{APPA} for more details), an effective non-Hermitian Hamiltonian is obtained as
\begin{eqnarray}\label{Eq9}
H^{\mathrm{eff}}_{\mathrm{ab}} &=& \omega _{a}a^{\dag }a+Ua^{\dag }a^{\dag }aa+\omega
_{b}b^{\dag }b  \nonumber \\
&&+ \left[ (J-iJ'e^{-i\phi})ab^{\dag }+ (J-iJ'e^{i\phi}) a^{\dag}b\right]
\end{eqnarray}
with the dissipation-induced coupling $J'=2G_{ac}G_{bc}/\gamma_c$ and the dissipation-induced decay rates $\gamma_a'=4G_{ac}^2/\gamma_c$ and $\gamma_b'=4G_{bc}^2/\gamma_c$.

The time-reversal symmetry of the Hamiltonian can be respected only with the synthetic magnetic flux $\phi = k\pi$ ($k$ is an integer). In order to observe clear nonreciprocal behaviors, we set $J'=2G_{ac}G_{bc}/\gamma_c=J$. When $\phi=\pi/2$, we have $J_{a\rightarrow b}\equiv J-iJ'e^{-i\phi}=0$ and $J_{b\rightarrow a}\equiv J-iJ'e^{i\phi}=2J$, so that photons can transmit from cavity $b$ to cavity $a$, but not vice versa. Moreover, the transmitted photons from cavity $b$ to cavity $a$ exhibit strong antibunching effect for strong nonlinear interaction in cavity $a$. Instead, if $\phi=3\pi/2$, we have $J_{a\rightarrow b}\equiv J-iJ'e^{-i\phi}=2J$ and $J_{b\rightarrow a}\equiv J-iJ'e^{i\phi}=0$, so that photons can only transmit from cavity $a$ to cavity $b$.

\begin{figure}[tbp]
\includegraphics[bb=199 315 400 508, width=8 cm, clip]{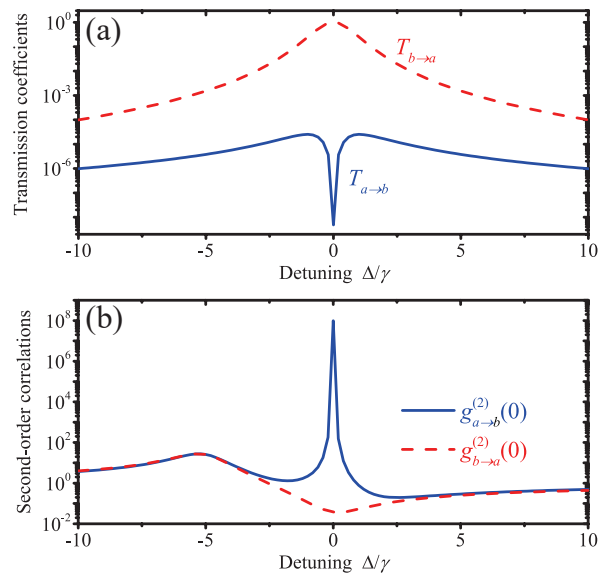}
\caption{(Color online) (a) The transmission coefficients $T_{a\rightarrow
b} $ (blue solid curve) and $T_{b\rightarrow a}$ (red dashed curve) are
plotted as functions of the detuning $\Delta/\protect\gamma$. (b) The
equal-time second-order correlation functions $g^{(2)}_{a%
\rightarrow b}(0)$ (blue solid curve) and $g^{(2)}_{b\rightarrow
a}(0)$ (red dashed curve) are plotted as functions of the detuning $\Delta/%
\protect\gamma$. Other used parameters are $\protect\gamma_a=\protect\gamma%
_b=\protect\gamma$, $\protect\gamma_c=100\gamma$, $\protect\varepsilon=0.01\protect\gamma$, $J=\protect%
\gamma/2$, $G_{ac}=G_{bc}=5\gamma$, $\phi=\pi/2$, and $U=5\protect\gamma$.}
\label{fig5}
\end{figure}

The analytical discussions are confirmed by numerically solving the master equation with Hamiltonian in Eq.~(\ref{Eq8}), using the parameters $\gamma_c= 100 \gamma$ and $G_{ac}=G_{bc}=5\gamma$ with $\gamma\equiv\gamma_a=\gamma_b$, as shown in Fig.~\ref{fig5}. Both the nonreciprocal transmission and the nonreciprocal photon blockade can be observed simultaneously via nonlinearity and reservoir engineering. Different from the results shown in Fig.~\ref{fig4}, we can observe nonreciprocal transmission over the full bandwidth, but nonreciprocal photon blockade only around the detuning $\Delta=0$.
We can also tune the nonreciprocity of the system by controlling the phase $\phi$, and very similar results can be obtained as shown in Figs.~\ref{fig4}(c) and \ref{fig4}(d).
In addition, the similar method can be applied to observe both the nonreciprocal transmission and the nonreciprocal photon blockade between two cavities without direct coupling, such as remote cavities or cavities with different frequencies, and the discussions are shown in Appendix~\ref{APPB}.

\section{Nonreciprocity in symmetric nonlinear optical molecule}

\begin{figure}[tbp]
\includegraphics[bb=194 322 395 599, width=8 cm, clip]{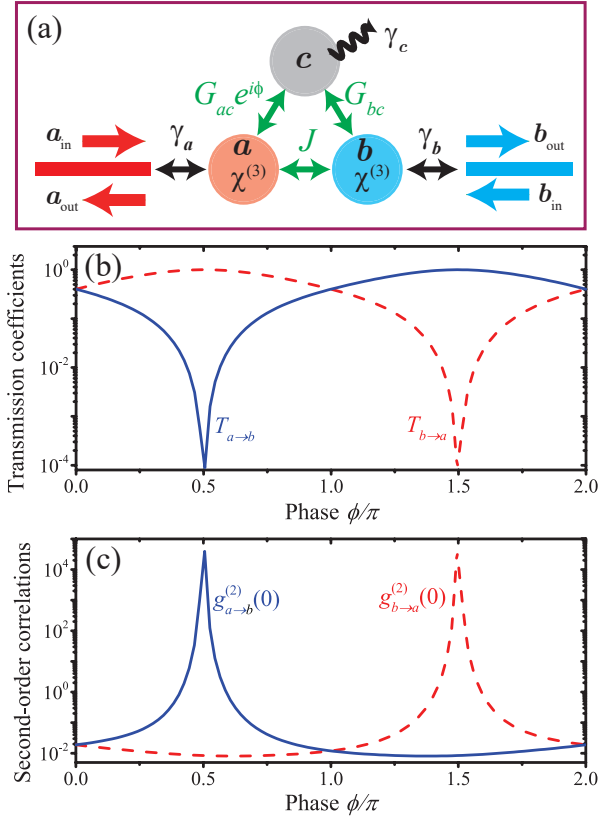}
\caption{(Color online) (a) Schematic diagram of a symmetric nonlinear optical molecule coupled to an auxiliary (mechanical or optical) mode $c$, with real coupling strengths ($J$, $G_{ac}$ and $G_{bc}$) and total phase $\phi$. (b) The transmission coefficients $T_{a\rightarrow
b} $ (blue solid curve) and $T_{b\rightarrow a}$ (red dashed curve) are
plotted as functions of the phase $\phi/\pi$. (c) The
equal-time second-order correlation functions $g^{(2)}_{a%
\rightarrow b}(0)$ (blue solid curve) and $g^{(2)}_{b\rightarrow
a}(0)$ (red dashed curve) are plotted as functions of the phase $\phi/\pi$. Other used parameters are $\Delta=0$, $\protect\gamma_a=\protect\gamma%
_b=\protect\gamma_c=\protect\gamma$, $\protect\varepsilon=0.01\protect\gamma$, $J=G_{ac}=G_{bc}=\protect%
\gamma/2$, and $U=5\protect\gamma$.}
\label{fig6}
\end{figure}

In the previous two sections, we have discussed the transmission
and the statistical properties of the transport photons in an asymmetric nonlinear optical molecule with the help of an auxiliary (mechanical or optical) mode $c$. While this configuration can realize strong nonreciprocal photon blockade in the direction from the linear cavity $b$ to the nonlinear cavity $a$, i.e., $g^{(2)}_{b\rightarrow
a}(0) \ll 1$, strong nonreciprocal blockade cannot be
realized in the direction from the nonlinear cavity $a$ to the linear cavity $b$, because the nonlinearity in cavity $a$ can not blockade all the transitions to the two-photon states in linear cavity $b$.
In order to realize strong photon blockade in both directions, we need a symmetric nonlinear optical molecule, i.e., both cavity $a$ and $b$ are filled with Kerr-type nonlinear material $\chi ^{(3)}$ (nonlinear interaction strength $U$).
However, different from the asymmetric nonlinear optical molecule discussed in Sec.~II, there is no nonreciprocity in the symmetric nonlinear optical molecule, i.e., $T_{b\rightarrow a}=T_{a\rightarrow b}$ and $g^{(2)}_{b\rightarrow a}(0)=g^{(2)}_{a\rightarrow b}(0)$, for the exchange symmetry between the two cavities.

In this section, we will show that nonreciprocal transmission and nonreciprocal photon blockade can be observed in both directions ($b \rightarrow a$ for $\phi=\pi/2$ and $a \rightarrow b$ for $\phi=3\pi/2$) with an auxiliary (mechanical or optical) mode $c$ coupling to a symmetric nonlinear optical molecule, as shown in Fig.~\ref{fig6}(a).
The Hamiltonian of the system can be written as
$H=H^{(2)}_{\mathrm{abc}}+H_{\mathrm{probe}}$, with
\begin{eqnarray}
H^{(2)}_{\mathrm{abc}} &=& \omega _{a}a^{\dag }a+Ua^{\dag }a^{\dag }aa+\omega
_{b}b^{\dag }b+Ub^{\dag }b^{\dag }bb+ \omega _{c}c^{\dag }c  \nonumber \\
&&+ \left( Jab^{\dag }+G_{bc}b c^{\dag }+G_{ac} e^{i \phi} c a^{\dag
}+\mathrm{H.c.}\right),
\end{eqnarray}
and $H_{\mathrm{probe}}$ in Eq.~(\ref{Eq2}).

The transmission coefficients are plotted as functions of the phase $\phi$ in Fig.~\ref{fig6}(b), which are almost the same as the results shown in Fig.~\ref{fig4}(c). This illustrates again the fact that the transmission coefficients are nearly independent of the nonlinearity of the system, because the populations in the two-photon states are much smaller than the ones in the one-photon states when the driving field is weak.
More importantly, the statistical properties of the transport photons shown in Fig.~\ref{fig6}(c) are different from the results shown in Fig.~\ref{fig4}(d). Firstly, we have $g^{(2)}_{b\rightarrow a}(0)<g^{(2)}_{a\rightarrow b}(0)$ with phase $0<\phi <\pi$ and $g^{(2)}_{b\rightarrow a}(0)>g^{(2)}_{a\rightarrow b}(0)$ with phase $\pi <\phi <2\pi$ in Fig.~\ref{fig6}(c), but we have $g^{(2)}_{b\rightarrow a}(0)<g^{(2)}_{a\rightarrow b}(0)$ for $0 \leq \phi \leq 2\pi$ in Fig.~\ref{fig4}(d). It shows that we can realize strong nonreciprocal photon blockade in both directions with different phase ($b \rightarrow a$ for $\phi=\pi/2$ and $a \rightarrow b$ for $\phi=3\pi/2$) with an auxiliary (mechanical or optical) mode $c$ coupling to a symmetric nonlinear optical molecule.

\section{Circulator with nonreciprocal photon blockade}

\begin{figure}[tbp]
\includegraphics[bb=194 325 395 650, width=8 cm, clip]{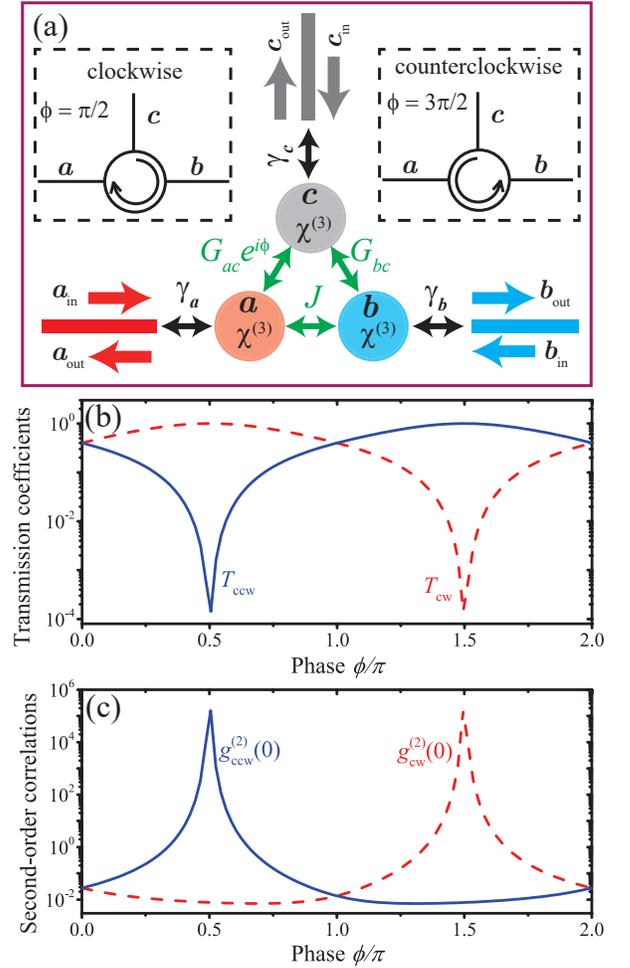}
\caption{(Color online) (a) Schematic diagram of symmetric nonlinear cyclic three-mode systems. (b) The transmission coefficients $T_{\rm ccw} $ (blue solid curve) and $T_{\rm cw}$ (red dashed curve) are
plotted as functions of the phase $\phi/\pi$. (c) The
equal-time second-order correlation functions $g^{(2)}_{\rm ccw}(0)$ (blue solid curve) and $g^{(2)}_{\rm cw}(0)$ (red dashed curve) are plotted as functions of the phase $\phi/\pi$. Other used parameters are $\Delta=0$, $\protect\gamma_a=\protect\gamma%
_b=\protect\gamma_c=\protect\gamma$, $\protect\varepsilon=0.01\protect\gamma$, $J=G_{ac}=G_{bc}=\protect%
\gamma/2$, and $U=5\protect\gamma$.}
\label{fig7}
\end{figure}

In this section, we present a circulator design based on an extension of the structure described in Section V, i.e., a symmetric nonlinear cyclic three-mode system, as shown in Fig.~\ref{fig7}(a).
The system is consisting of three mutually coupled cavities which are filled with Kerr-type nonlinear material $\chi ^{(3)}$ (nonlinear interaction strength $U$) respectively.
The Hamiltonian of the system can be written as
$H=H^{(3)}_{\mathrm{abc}}+H_{\mathrm{probe}}$, with
\begin{eqnarray}
H^{(3)}_{\mathrm{abc}} &=& \omega _{a}a^{\dag }a+\omega_{b}b^{\dag }b+ \omega _{c}c^{\dag }c  \nonumber \\
&&+Ua^{\dag }a^{\dag }aa+Ub^{\dag }b^{\dag }bb+Uc^{\dag }c^{\dag }cc  \nonumber \\
&&+ \left( Jab^{\dag }+G_{bc}b c^{\dag }+G_{ac} e^{i \phi} c a^{\dag
}+\mathrm{H.c.}\right),
\end{eqnarray}
and $H_{\mathrm{probe}}$ in Eq.~(\ref{Eq2}) for $o=a,b,c$.

When photons are input from cavity $c$, then we have $c_{%
\mathrm{in}}=\varepsilon /\sqrt{\gamma _{c}}$ and the transmission coefficients
from cavity $c$ to cavity $a$ and cavity $b$ can be defined by
\begin{equation}
T_{c\rightarrow a}\equiv \frac{\langle a_{\mathrm{out}}^{\dag }a_{\mathrm{out%
}}\rangle }{\langle c_{\mathrm{in}}^{\dag }c_{\mathrm{in}}\rangle }=\frac{%
\gamma _{a}\gamma _{c}}{\varepsilon ^{2}}\left\langle a^{\dag }a\right\rangle,
\end{equation}
\begin{equation}
T_{c\rightarrow b}\equiv \frac{\langle b_{\mathrm{out}}^{\dag }b_{\mathrm{out%
}}\rangle }{\langle c_{\mathrm{in}}^{\dag }c_{\mathrm{in}}\rangle }=\frac{%
\gamma _{b}\gamma _{c}}{\varepsilon ^{2}}\left\langle b^{\dag }b\right\rangle.
\end{equation}
with the statistic properties of the transmitted photons described by
\begin{equation}
g_{c\rightarrow a}^{(2)}(0)\equiv \frac{\langle a_{\mathrm{out}}^{\dag }a_{%
\mathrm{out}}^{\dag }a_{\mathrm{out}}a_{\mathrm{out}}\rangle }{\langle a_{%
\mathrm{out}}^{\dag }a_{\mathrm{out}}\rangle ^{2}}=\frac{\langle a^{\dag
}a^{\dag }aa\rangle }{\langle a^{\dag }a\rangle ^{2}},
\end{equation}
\begin{equation}
g_{c\rightarrow b}^{(2)}(0)\equiv \frac{\langle b_{\mathrm{out}}^{\dag }b_{%
\mathrm{out}}^{\dag }b_{\mathrm{out}}b_{\mathrm{out}}\rangle }{\langle b_{%
\mathrm{out}}^{\dag }b_{\mathrm{out}}\rangle ^{2}}=\frac{\langle b^{\dag
}b^{\dag }bb\rangle }{\langle b^{\dag }b\rangle ^{2}}.
\end{equation}
Similarly, the transmission coefficient from cavity $a$ (cavity $b$) to cavity $c$ with $c_{\mathrm{out}}=\sqrt{\gamma _{c}}c$ can be defined by
\begin{equation}
T_{a\rightarrow c}\equiv \frac{\langle c_{\mathrm{out}}^{\dag }c_{\mathrm{out%
}}\rangle }{\langle a_{\mathrm{in}}^{\dag }a_{\mathrm{in}}\rangle }=\frac{%
\gamma _{a}\gamma _{c}}{\varepsilon ^{2}}\left\langle c^{\dag }c\right\rangle,
\end{equation}
\begin{equation}
T_{b\rightarrow c}\equiv \frac{\langle c_{\mathrm{out}}^{\dag }c_{\mathrm{out%
}}\rangle }{\langle b_{\mathrm{in}}^{\dag }b_{\mathrm{in}}\rangle }=\frac{%
\gamma _{b}\gamma _{c}}{\varepsilon ^{2}}\left\langle c^{\dag }c\right\rangle,
\end{equation}
and the statistic properties of the transmitted photons can be described by the equal-time second-order
correlation functions in the steady state ($t\rightarrow \infty $) as
\begin{equation}
g_{a/b\rightarrow c}^{(2)}(0)\equiv \frac{\langle c_{\mathrm{out}}^{\dag }c_{%
\mathrm{out}}^{\dag }c_{\mathrm{out}}c_{\mathrm{out}}\rangle }{\langle c_{%
\mathrm{out}}^{\dag }c_{\mathrm{out}}\rangle ^{2}}=\frac{\langle c^{\dag
}c^{\dag }cc\rangle }{\langle c^{\dag }c\rangle ^{2}}.
\end{equation}%
With permutation symmetry in the nonlinear cyclic three-mode systems for $\protect\omega_a=\protect\omega%
_b=\protect\omega_c$, $\protect\gamma_a=\protect\gamma%
_b=\protect\gamma_c=\gamma$, and $J=G_{ac}=G_{bc}=\protect\gamma/2$, we have $T_{\rm ccw}\equiv T_{a\rightarrow b}=T_{b\rightarrow c}=T_{c\rightarrow a}$, $T_{\rm cw}\equiv T_{a\rightarrow c}=T_{c\rightarrow b}=T_{b\rightarrow a}$, $g^{(2)}_{\rm ccw}(0)\equiv g^{(2)}_{a\rightarrow b}(0)=g^{(2)}_{b\rightarrow c}(0)=g^{(2)}_{c\rightarrow a}(0)$, and  $g^{(2)}_{\rm cw}(0)\equiv g^{(2)}_{a\rightarrow c}(0)=g^{(2)}_{c\rightarrow b}(0)=g^{(2)}_{b\rightarrow a}(0)$.

The transmission coefficients and second-order correlation functions are plotted as functions of the phase $\phi$ in Figs.~\ref{fig7}(b) and \ref{fig7}(c) under the resonance condition $\Delta=0$. The photons transport clockwise ($T_{\rm cw}\approx1$, $T_{\rm ccw}\approx 0$) with nonreciprocal photon blockade ($g^{(2)}_{\rm cw}(0)\ll 1 $, $g^{(2)}_{\rm ccw}(0)\gg 1 $) with phase $\phi=\pi/2$, or transport counterclockwise ($T_{\rm ccw}\approx1$, $T_{\rm cw}\approx 0$) with nonreciprocal photon blockade ($g^{(2)}_{\rm ccw}(0)\ll 1 $ and $g^{(2)}_{\rm cw}(0)\gg 1 $) with phase $\phi=3\pi/2$. The symmetric nonlinear cyclic three-mode system provides us a platform to realize a circulator with nonreciprocal photon blockade, and may have important applications in quantum information processing.

\section{Conclusions}

In conclusion, we have shown that both nonreciprocal transmission and nonreciprocal photon blockade can be observed in a nonlinear optical molecule via nonlinearity and synthetic magnetism, which can serve as a nonreciprocal single-photon source to create single photons in a desired direction, or manipulate one-way nonclassical light as a single-photon diode.
Moreover, a circulator with nonreciprocal photon blockade was designed based on the combination of nonlinearity and synthetic magnetism in a symmetric nonlinear cyclic three-mode system.
The nonreciprocity based on the combination of nonlinearity and synthetic magnetism can also be used for other applications, such
as nonreciprocal photon turnstiles~\cite{DayanSci}, nonreciprocal photon routers~\cite{AokiPRL09,LZhou13,ShomroniSci14}, and directional amplifiers~\cite{YLiOE17,CJiangPRA18,XZZhangPRA18,MalzPRL18,LepinayPRAPP19}.

The combination of synthetic magnetism and nonlinearity is a general method to show both nonreciprocal transmission and nonreciprocal photon blockade simultaneously, and could be implemented in photonic systems, microwave superconducting circuits, and optomechanical systems~\cite{LTianPRA17,GLiPRA18}.
Our work can also be extended to a wide range of systems with the Kerr nonlinearity replaced by second-order nonlinearity~\cite{XZhangNPo19}, optomechanical interaction~\cite{ZWangSR15,XuXWPRA18,LNSongArx19}, and the interaction to a two-level quantum emitter~\cite{ASZhengSR17} or two-level atomic ensemble~\cite{LNSongOC18}.

\vskip 2pc \leftline{\bf Acknowledgement}

X.-W.X. was supported by the National Natural Science Foundation of China
(NSFC) under Grant No.~11604096, and the Key Program of Natural Science Foundation of Jiangxi Province, China under Grant No.~20192ACB21002. A.-X.C. is supported by NSFC under Grant
No.~11775190. Y.L. is supported by NSFC under Grant
No.~11774024.  H.J. is supported by NSFC under Grants No.~11474087 and No.~11774086.

\appendix

\section{The derivation of the effective Hamiltonian in Eq.~(\ref{Eq9})}\label{APPA}

The effective Hamiltonian given in Eq.~(\ref{Eq9}) can be obtained from Eq.~(\ref{Eq8}) by adiabatically eliminating the auxiliary mode $c$ in the formalism of quantum Langevin equations, by using the conditions $\gamma_c\gg \max\{\gamma_a, \gamma_b,J,G_{ac},G_{bc}\}$ and $\omega_a=\omega_b=\omega_c$.
In the rotating frame at the resonance frequency of the modes $\omega_a=\omega_b=\omega_c$, the Hamiltonian in Eq.~(\ref{Eq8}) can be rewritten as
\begin{equation}
\widetilde{H}^{(1)}_{\mathrm{abc}}=Ua^{\dag }a^{\dag }aa+\left( Jab^{\dag
}+G_{bc}bc^{\dag }+G_{ac}e^{i\phi }ca^{\dag }+\mathrm{H.c.}\right).
\end{equation}%
The quantum Langevin equations for the annihilation operators are given by
\begin{eqnarray}
\frac{d}{dt}a &=&-\frac{\gamma _{a}}{2}a-i2Ua^{\dag }aa-iG_{ac}e^{i\phi }c \nonumber\\
&&-iJb+\sqrt{\gamma _{a}}a_{\rm in},
\end{eqnarray}%
\begin{equation}
\frac{d}{dt}b=-\frac{\gamma _{b}}{2}b-iJa-iG_{bc}c+\sqrt{\gamma _{b}}b_{\rm in},
\end{equation}
\begin{equation}
\frac{d}{dt}c=-\frac{\gamma _{c}}{2}c-iG_{bc}b-iG_{ac}e^{-i\phi }a+\sqrt{%
\gamma _{c}}c_{\rm in},
\end{equation}
where $a_{\rm in}$, $b_{\rm in}$, and $c_{\rm in}$ are the input fields, including the noises from the environments.
Under the adiabatical condition $\gamma_c\gg \max\{\gamma_a=\gamma_b,J,G_{ac},G_{bc}\}$, the annihilation operator $c$ can be given approximately as
\begin{equation}
c=-i\frac{2}{\gamma _{c}}G_{bc}b-i\frac{2}{\gamma _{c}}G_{ac}e^{-i\phi }a+C_{%
\mathrm{noise}},
\end{equation}
where $C_{\mathrm{noise}}$ is the noise coming from the environments as there is no single input from auxiliary mode $c$ in this model.
Substituting the above relation into dynamical equations of $a$ and $b$, then the effective dynamical equations of $a$ and $b$ are given by
\begin{eqnarray}
\frac{d}{dt}a &=&-\frac{1}{2}\left( \gamma _{a}+\gamma _{a}^{\prime }\right)
a-i2Ua^{\dag }aa-i\left( J-iJ^{\prime }e^{i\phi }\right) b \nonumber\\
&&+\sqrt{\gamma _{a}}%
a_{in}-iG_{ac}e^{i\phi }C_{\mathrm{noise}}
\end{eqnarray}%
\begin{eqnarray}
\frac{d}{dt}b &=&-\frac{1}{2}\left( \gamma _{b}+\gamma _{b}^{\prime }\right)
b-i\left( J-iJ^{\prime }e^{-i\phi }\right) a  \nonumber\\
&&+\sqrt{\gamma _{b}}b_{in}-iG_{bc}C_{\mathrm{noise}}
\end{eqnarray}%
where $J^{\prime }\equiv 2G_{ac}G_{bc}/\gamma _{c}$ is the coupling strength, and $\gamma _{b}^{\prime
}\equiv 2G_{bc}^{2}/\gamma _{c}$ and $\gamma _{a}^{\prime }\equiv
4G_{ac}^{2}/\gamma $ are the decay rates, induced by the auxiliary mode $c$. These dynamical equations can also be derived from the following effective Hamiltonian by adding dissipations,
\begin{equation}
\widetilde{H}_{\mathrm{ab}}^{\mathrm{eff}}=Ua^{\dag }a^{\dag }aa+\left(
J-iJ^{\prime }e^{-i\phi }\right) ab^{\dag }+\left( J-iJ^{\prime }e^{i\phi
}\right) a^{\dag }b,
\end{equation}
which is the effective Hamiltonian given in Eq.~(\ref{Eq9}) in the rotating frame at the resonance frequency of the modes $\omega_a=\omega_b$.

\section{Nonreciprocity without direct coupling}\label{APPB}

\begin{figure}[tbp]
\includegraphics[bb=194 324 391 635, width=8 cm, clip]{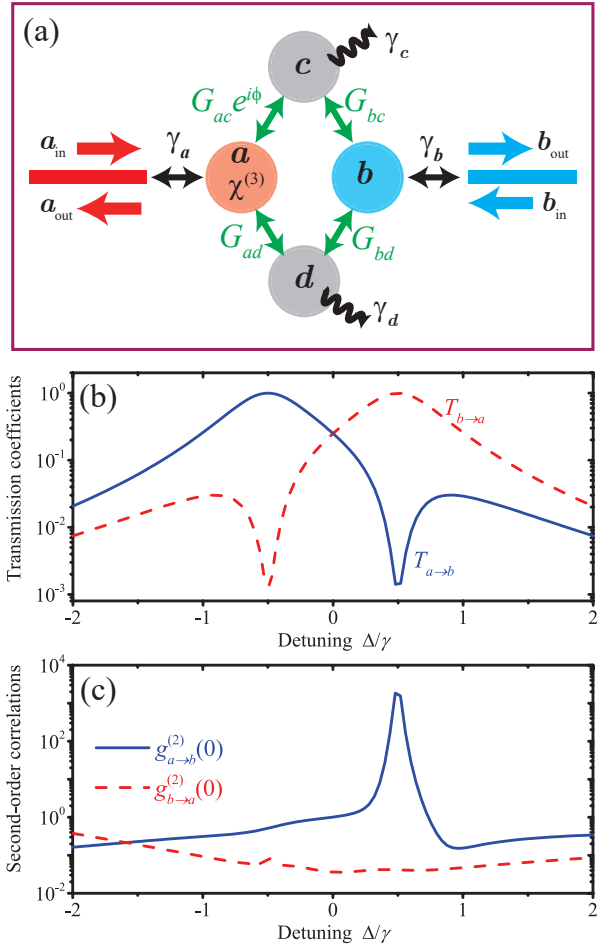}
\caption{(Color online) (a) Schematic diagram of two optical modes (nonlinear
mode $a$ and linear mode $b$) coupled indirectly by two auxiliary (mechanical or optical) modes ($c$ and $d$) simultaneously, with real coupling strengths $G_{ac}$, $G_{ad}$, $G_{bc}$ and $G_{bd}$, and total phase $\phi$. (b) The transmission coefficients $T_{a\rightarrow
b} $ (blue solid curve) and $T_{b\rightarrow a}$ (red dashed curve) are
plotted as functions of the detuning $\Delta/\protect\gamma$. (c) The
equal-time second-order correlation functions $g^{(2)}_{a%
\rightarrow b}(0)$ (blue solid curve) and $g^{(2)}_{b\rightarrow
a}(0)$ (red dashed curve) are plotted as functions of the detuning $\Delta/%
\protect\gamma$. Other used parameters are $\protect\gamma_a=\protect\gamma%
_b=\protect\gamma$, $\protect\gamma_c=\gamma/1000$, $\protect\gamma_d=16\gamma$, $\protect\varepsilon=0.01\protect\gamma$, $G_{ac}=G_{bc}=\gamma/2$, $G_{ad}=G_{bd}=2\gamma$, $\phi=\pi/2$, and $U=5\protect\gamma$.}
\label{fig8}
\end{figure}

Both the nonreciprocal transmission and the nonreciprocal photon blockade can also be observed between two cavities without direct coupling, such as remote cavities or cavities with different frequencies. As shown in Fig.~\ref{fig8}(a),
two cavities ($a$ and $b$) are coupled indirectly by two auxiliary (mechanical or optical) modes ($c$ and $d$, frequencies $\omega_c$ and $\omega_d$) with real coupling strengths $G_{ac}$, $G_{ad}$, $G_{bc}$ and $G_{bd}$, and total phase $\phi$. A similar model without nonlinearity has been investigated in Refs.~\cite{XuXWPRA16,PetersonPRX17,BernierNC17,BarzanjehNC17}, and the nonreciprocal transmission between cavities $a$ and $b$ has been predicted theoretically~\cite{XuXWPRA16} and demonstrated experimentally~\cite{PetersonPRX17,BernierNC17,BarzanjehNC17}.
Here, we will show that this model can exhibit both nonreciprocal transmission and nonreciprocal photon blockade after introducing nonlinearity.

The model in Fig.~\ref{fig8}(a) can be described by the Hamiltonian $H=H^{(1)}_{\mathrm{abcd}}+H_{\mathrm{probe}}$, with
\begin{eqnarray}
H^{(1)}_{\mathrm{abcd}} &=& \omega _{a}a^{\dag }a+Ua^{\dag }a^{\dag }aa+\omega
_{b}b^{\dag }b+ \omega _{c}c^{\dag }c+ \omega _{d}d^{\dag }d \nonumber\\
&&+ \left( G_{ad}ad^{\dag }+G_{ac} e^{i \phi} c a^{\dag} \right.  \nonumber  \\
&& \left.+ G_{bc}b c^{\dag }+G_{bd}d b^{\dag }+\mathrm{H.c.}\right),
\end{eqnarray}
and $H_{\mathrm{probe}}$ in Eq.~(\ref{Eq2}).
In the numerical calculation of the transmission coefficients and second-order correlation functions, we set the damping rates of the auxiliary modes ($c$ and $d$) as $\gamma _{c}$ and $\gamma _{d}$, and the dissipation of the auxiliary modes, i.e., $\gamma _{c} L[c]\rho+\gamma _{d} L[d]\rho$, is added to the master equation.
We assume that all the modes shall the same frequency, i.e., $\omega _{a}=\omega _{b}=\omega _{c}=\omega _{d}$, and the decay rates ($\gamma_c$ and $\gamma_d$) of the two auxiliary modes ($c$ and $d$) satisfy the condition
\begin{equation}
\gamma_d \gg \min\{G_{ij},\gamma_a,\gamma_b\} \gg \gamma_c,
\end{equation}
where $i=a,b$ and $j=c,d$.

The nonreciprocal transmission of photons in the cyclic four-mode system [in Fig.~\ref{fig8}(a)]
is induced by breaking the time-reversal symmetry of the system with the synthetic magnetic flux $\phi \neq k\pi$ ($k$ is an integer). Physically, the transport photons from one cavity to the other one undergo
a Mach-Zehnder-type interference: one path is the hopping
through the auxiliary mode $c$ and the other path is the
hopping through the auxiliary mode $d$. In the time-reversal symmetry broken regime, i.e., $\phi \neq k\pi$, the photons transport in one direction undergo a constructive interference, and the transmission coefficient is enhanced. Meanwhile, the transmission coefficient in the reverse direction is suppressed with destructive interference.
As shown in Ref.~\cite{XuXWPRA16} without nonlinearity ($U=0$), the perfect nonreciprocal transmission i.e., $T_{a \rightarrow b}=1$ and $T_{b \rightarrow a}=0$ or $T_{a \rightarrow b}=0$ and $T_{b \rightarrow a}=1$, is obtained with the parameters $G_{ac}=G_{bc}=\gamma/2$, $G_{ad}=G_{bd}=\sqrt{\gamma\gamma_d}/2$, $\phi=\pm\pi/2$, and detuning $\Delta \equiv \omega-\omega_p  = \pm \gamma/2 $, where $\gamma\equiv \gamma_a=\gamma_b$ and $\omega\equiv \omega_a=\omega_b=\omega_c=\omega_d$.
With the same parameters, the similar results are demonstrated numerically in Fig.~\ref{fig8}(b) for $\phi=\pi/2$ and $U=5\protect\gamma$.
We have $T_{a \rightarrow b}=1$ and $T_{b \rightarrow a}=0$ around the detuning $\Delta = - \gamma/2 $, and $T_{a \rightarrow b}=0$ and $T_{b \rightarrow a}=1$ for detuning $\Delta \approx \gamma/2 $.

\begin{figure}[tbp]
\includegraphics[bb=194 323 392 635, width=8 cm, clip]{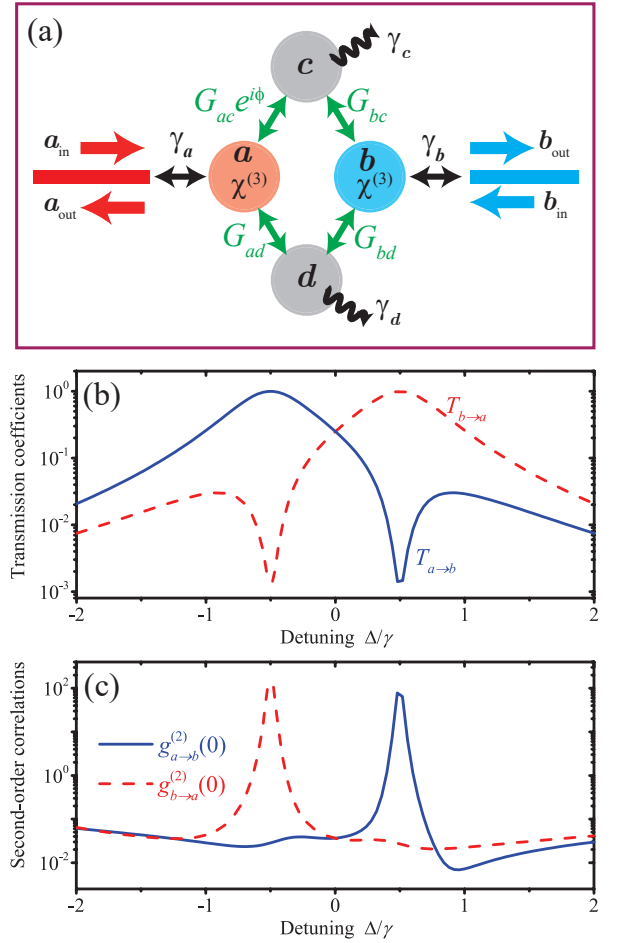}
\caption{(Color online) (a) Schematic diagram of two optical modes (nonlinear
mode $a$ and nonlinear mode $b$) coupled indirectly by two auxiliary (mechanical or optical) modes ($c$ and $d$) simultaneously, with real coupling strengths $G_{ac}$, $G_{ad}$, $G_{bc}$ and $G_{bd}$, and total phase $\phi$. (b) The transmission coefficients $T_{a\rightarrow
b} $ (blue solid curve) and $T_{b\rightarrow a}$ (red dashed curve) are
plotted as functions of the detuning $\Delta/\protect\gamma$. (c) The
equal-time second-order correlation functions $g^{(2)}_{a%
\rightarrow b}(0)$ (blue solid curve) and $g^{(2)}_{b\rightarrow
a}(0)$ (red dashed curve) are plotted as functions of the detuning $\Delta/%
\protect\gamma$. Other used parameters are $\protect\gamma_a=\protect\gamma%
_b=\protect\gamma$, $\protect\gamma_c=\gamma/1000$, $\protect\gamma_d=16\gamma$, $\protect\varepsilon=0.01\protect\gamma$, $G_{ac}=G_{bc}=\gamma/2$, $G_{ad}=G_{bd}=2\gamma$, $\phi=\pi/2$, and $U=5\protect\gamma$.}
\label{fig9}
\end{figure}

As shown in Fig.~\ref{fig8}(c), nonreciprocal photon blockade, i.e., $g^{(2)}_{b\rightarrow a}(0)\ll 1$ and $g^{(2)}_{a\rightarrow b}(0)\gg 1$, can be observed around the detuning $\Delta =  \gamma/2 $ for phase $\phi = \pi/2 $, which are corresponding to the conditions for $T_{a \rightarrow b}=0$ and $T_{b \rightarrow a}=1$.
In other words, the photons transport one by one with high transmission coefficients from cavity $a$ to cavity $b$, but transport in pairs with low transmission coefficients in the reverse direction.
As already mentioned in the above sections, the photons transport one by one is induced by the strong nonlinear interaction in cavity $a$, and the photons transport in pairs are generated by quenching the population of the one-photon state in cavity $b$, which is origin from the destructive interference~\cite{XuXWPRA14} between the two paths for generating one photon in cavity $b$.
In addition, nonreciprocal photon blockade with $g^{(2)}_{b\rightarrow a}(0)\ll 1$ and $g^{(2)}_{a\rightarrow b}(0)\gg 1$, and nonreciprocal transmission with $T_{a \rightarrow b}=0$ and $T_{b \rightarrow a}=1$, can also be observed around the detuning $\Delta =  -\gamma/2 $ for phase $\phi = - \pi/2 $, which are not shown here.

While the configuration shown in Fig.~\ref{fig8}(a) can realize strong nonreciprocal photon blockade in the direction from the linear cavity $b$ to the nonlinear cavity $a$, i.e., $g^{(2)}_{b\rightarrow a}(0) \ll 1$, around detuning $\Delta =\gamma/2$, strong nonreciprocal blockade cannot be realized in the direction from the nonlinear cavity $a$ to the linear cavity $b$ for detuning $\Delta =-\gamma/2$, because the nonlinearity in cavity $a$ can not blockade all the transitions to the two-photon states in linear cavity $b$.
In order to realize strong photon blockade in both directions, we can just replace the linear cavity $b$ by a nonlinear cavity $b$ as shown in Fig.~\ref{fig9}(a). Then the Hamiltonian $H^{(1)}_{\mathrm{abcd}}$ is replaced by
\begin{eqnarray}
H^{(2)}_{\mathrm{abcd}} &=& \omega _{a}a^{\dag }a+\omega
_{b}b^{\dag }b+ \omega _{c}c^{\dag }c+ \omega _{d}d^{\dag }d \nonumber\\
&&+Ua^{\dag }a^{\dag }aa+Ub^{\dag }b^{\dag }bb\nonumber\\
&&+ \left( G_{ad}ad^{\dag }+G_{ac} e^{i \phi} c a^{\dag} \right.  \nonumber  \\
&& \left.+ G_{bc}b c^{\dag }+G_{bd}d b^{\dag }+\mathrm{H.c.}\right).
\end{eqnarray}

The transmission coefficients and second-order correlation functions are shown in Figs.~\ref{fig9}(c) and \ref{fig9}(d), respectively. Clearly, the additional nonlinearity in cavity $b$ has no noticeable effect to the transmission coefficients. But it has significant effect on the statistical properties of the transmitted photons. In this case, strong nonreciprocal blockade, i.e., $g^{(2)}_{a\rightarrow b}(0)\ll 1$ and $g^{(2)}_{b\rightarrow a}(0)\gg 1$, can also be realized in the direction from cavity $a$ to cavity $b$ with high isolation rate ($T_{a \rightarrow b}\approx 1$ and $T_{b \rightarrow a}\approx 0$) around the detuning $\Delta =-\gamma/2$. Thus we can design a quantum nonreciprocal device for two weak signals at different frequencies $\Delta =\pm\gamma/2$ in reverse directions at the same time.

\bibliographystyle{apsrev}

\end{document}